\documentclass[a4paper, amsfonts, amssymb, amsmath, reprint, showkeys, nofootinbib, twoside]{revtex4-1}

\usepackage[english]{babel}
\usepackage[utf8]{inputenc}
\usepackage[dvipsnames]{xcolor}
\usepackage[colorinlistoftodos, color=green!40, prependcaption]{todonotes}
\usepackage{amsthm}
\usepackage{mathtools}
\usepackage{physics}
\usepackage[left=23mm,right=13mm,top=35mm,columnsep=15pt]{geometry} 
\usepackage{adjustbox}
\usepackage{placeins}
\usepackage[T1]{fontenc}
\usepackage{lipsum}
\usepackage{csquotes}
\usepackage[pdftex, pdftitle={Article}, pdfauthor={Author}]{hyperref} 
\usepackage{graphicx}
\usepackage{dcolumn}
\usepackage{bm}
\usepackage[font=small,labelfont=bf, labelsep=endash]{caption}
\usepackage{subcaption}
\usepackage{ulem} 
\usepackage{soul}
\newcommand{\heading}[1]{\vspace{0.25truecm}\hspace{0.25truecm}\emph{\textbf{#1 ---}}}

\newcommand{\bA}{\mathbf{A}}

\newcommand{\bP}{\mathbf{P}}
\renewcommand{\pv}{\textit{p-value~}}
\newcommand{\pvs}{\textit{p-values~}}

\newcommand{\nchoosek}[2]{\left( \begin{array}{c} #1 \\ #2 \end{array} \right)}
\newcommand{\lfig}[1]{Fig.~\ref{#1}}
\renewcommand{\eqref}[1]{Eq.~(\ref{#1})}

\begin{document}

\title{Measuring topological descriptors of complex networks under uncertainty}

\author{Sebastian Raimondo}
    \email{sraimondo@fbk.eu}
    \affiliation{CoMuNe Lab, Center for Information and Communication Technology,
    Fondazione Bruno Kessler, Via Sommarive 18, 38123 Povo (TN), Italy.}
    \affiliation{Department of Mathematics, University of Trento, Via Sommarive 9, 38123 Povo (TN), Italy}
\author{Manlio De Domenico}
    \email{mdedomenico@fbk.eu}
    \affiliation{CoMuNe Lab, Center for Information and Communication Technology,
Fondazione Bruno Kessler, Via Sommarive 18, 38123 Povo (TN), Italy.}

\date{\today} 

\begin{abstract}
Revealing the structural features of a complex system from the observed collective dynamics is a fundamental problem in network science. In order to compute the various topological descriptors commonly used to characterize the structure of a complex system (e.g. the degree, the clustering coefficient), it is usually necessary to completely reconstruct the network of relations between the subsystems.
Several methods are available to detect the existence of interactions between the nodes of a network. By observing some physical quantities through time, the structural relationships are inferred using various discriminating statistics (e.g. correlations, mutual information, etc.). In this setting, the uncertainty about the existence of the edges is reflected in the uncertainty about the topological descriptors.

In this study, we propose a novel methodological framework to evaluate this uncertainty, replacing the topological descriptors, even at the level of a single node, with appropriate probability distributions, eluding the reconstruction phase.
Our theoretical framework agrees with the numerical experiments performed on a large set of synthetic and real-world networks. 
Our results provide a grounded framework for the analysis and the interpretation of widely used topological descriptors, such as degree centrality, clustering and clusters, in scenarios where the existence of network connectivity is statistically inferred or when the probabilities of existence $\pi_{ij}$ of the edges are known. To this purpose we also provide a simple and mathematically grounded process to transform the discriminating statistics into the probabilities $\pi_{ij}$.

\end{abstract}


\maketitle

\section{Introduction}

Complex natural and artificial systems are composed of many interacting dynamical units which exhibit a collective behavior~\cite{newman2003structure}. This is the result of the interplay between the dynamics of the constituents and the interactions among them. The structure of the interactions and the (nonlinear) dynamics have to be considered simultaneously to model such systems~\cite{boccaletti2006complex}. Unfortunately, the structure of many empirical systems usually remains hidden, but the dynamics of some physical quantity can be observed and measured.  From such observations the connectivity can be inferred for a broad class of systems \cite{yang, zou2019complex}, from the human brain \cite{manlio2016, chavez, pernice2011structure, bassett2017network}, to financial \cite{kumar2012correlation, bonanno2003topology}, weather and climate systems \cite{tsonis2008topology, zhou2015teleconnection}, including hydrological processes \cite{sivakumar2015network,boers} and biological systems \cite{albert2007network, hecker2009gene}. 

Many topological descriptors are used to characterize the structural features of a complex system (e.g. the degree, the transitivity, etc.), but to compute them an earlier reconstruction of the structure itself is usually necessary. The goal of network reconstruction is typically to solve this inverse problem~\cite{casadiego_timme}: from information about the dynamics, reconstruct the network of interactions. 
In general, a complex system can be described as follows: let $\mathbf{x_i}(t)$ denote the internal $D$-dimensional state $\mathbf{x_i}(t)=[x_i^{(1)},x_i^{(2)},\dots,x_i^{(D)}]^{\operatorname{T}}$ of a system consisting of $N$ dynamical units, at time $t$. The evolution of the state is governed by the system of $N$ ordinary differential equations
\begin{eqnarray*}
	\dot{\mathbf{x}}_i(t) = \Psi_i(\mathbf{x_i}(t),\boldsymbol{\gamma}_i) + \sum_{j=1}^{N}{A_{ij} \Phi_{ij}( \mathbf{x_i,x_j} ) } + \mathbf{u_i}(t) + \boldsymbol{\eta_i}(t)
\end{eqnarray*}
where $i,j \in \{1,2,\dots,N\}$, $ t \in \mathbb{R}$ ; the function $\Psi_i :  \mathbb{R}^D \to  \mathbb{R}^D $ and  $\Phi_{ij} :  \mathbb{R}^D \times  \mathbb{R}^D \to \mathbb{R}^D  $ respectively define the  intrinsic and interaction dynamics of the $D$-dimensional units. The function $\mathbf{u}(t)$ represents external drivers, $\boldsymbol{\eta}(t)$ is a dynamic noise term and $\boldsymbol{\gamma_i}$ is a set of dynamic parameters. Finally, the term $A_{ij}$ defines the interaction topology in terms of the adjacency matrix $\bA$ such that $A_{ij} = 1$ if there is a direct physical interaction from unit $j$ to $i$ and $A_{ij} = 0$ otherwise. This matrix completely defines a network, that is, an abstraction used to model a system that contains discrete, interconnected elements. The elements are represented by nodes (also called vertices) and the interconnections are represented by edges. In general, one should take into account the response of the experimental setup used for measuring the state (and the measurement noise), resulting in a vector $s(\mathbf{x(t)})$ of measured observables which is a function of $\mathbf{x}(t)$. In many cases, the reconstruction problem relies solely on the vector $s(\mathbf{x(t)})$, a multivariate time series. 
Many different methods have been proposed to recover the structure of the interactions between dynamical units from time series~(see e.g.~\cite{nitzan2017revealing, lacasa, runge, runge2012, runge2019inferring, runge_kretschmer2016, mccracken2014convergent, timme2014revealing, kumar2012correlation, zanin2018topological, timme}). The most widely used in practice, consist in the quantification of the interaction between units through an appropriate discriminating statistic -- typically measuring pairwise correlations~\cite{bassett_cor} or statistical causality between units~\cite{sun}, or mapping information flow from the observed collective dynamics~\cite{runge_multi} -- and then to apply a criterion to decide whether the measured interaction is significant or not~\cite{bassett2014cross, saint2020bionetwork, pvalue_thresh_bullmore}. The choice of the criterion is crucial, but typically it introduces some arbitrary choices in the process. The current reconstruction procedures often rely on heuristics to choose a threshold value for the pairwise correlation or causality measures. Values below the threshold are discarded, so that an edge is assigned only between units whose interaction is sufficiently strong. This procedure is known to produce complex features even when no complex structure is present~\cite{cantwell}. Preferably, using a more sophisticated statistical analysis, a set of \pvs is computed to evaluate the significance of the edge between the nodes with respect to a null-model~\cite{nakamura}. However, even in this case, the process incurs in the issues of partial correlations \cite{runge} and multiple testing~\cite{dudoit2003multiple}. Other approaches have been recently proposed (e.g. \cite{tiago, young2020}), by which the posterior probability distribution of the network structure is computed using suitable generative processes and prior information. The network is reconstructed by sampling from this distribution.  These approaches require the model of the dynamics to be defined together with its corresponding probabilistic model for the data. In other cases, the network structure can be constructed from static observations and ad-hoc measurements (see e.g. \cite{protein_net, static_reconstruction_sensors, road_net}), which may be affected by noise and measurement error. In any case, after the reconstruction of the network structure, it is possible to compute the topological descriptors of the structure.

In this study, we propose a new methodological framework to analyse  the structural features of a complex network when its topological connectivity is specified by edge probabilities, without the explicit reconstruction of the network structure. In addition, we propose a simple procedure to obtain the edge probabilities, given the \pvs that quantify the supporting evidence of the related discriminating statistics. The network descriptors are redefined as stochastic variables, whose probability distributions can be used to infer the relevant statistics and to evaluate their robustness against the uncertainty.
Note that this framework complements other approaches like \cite{letopeel} -- which regards community detection -- and \cite{tiago, young2020} -- where a specific generative model for the data is used. In fact, our approach aims at computing the topological descriptors of a complex network having information about the edge existence, without reconstructing the entire network structure and in the absence of a model for the dynamical process.
Moreover, the proposed method does not build on assumptions about the topological features of the underlying network, nor on its generative process, but rather includes the prior knowledge about the existence of each edge. Employing a Bayesian procedure we derive for every $i$ and $j$ the probability $\pi_{ij}$ that the node $i$ is linked to the node $j$, given the \pv from the above mentioned analyses. Hence, the actual complex network is considered as a realization from the possibilities encoded in the probabilistic model that we call \enquote{fuzzy network} model.
Under this probabilistic perspective, all the network descriptors must be redefined as random variables. A natural way to recover the descriptive information is to consider the whole distribution or a suitable statistic. Therefore, we have defined the ``fuzzy'' counterpart of some basic structural descriptors such as the node degree and the network expected degree, the clustering coefficient, and the probability of having a unique connected component. For each of them, we present the analytical probability distributions and the main statistics. We applied this framework to various well-known synthetic and real-world networks starting from multivariate time series, and compared the results to the ones from a classical reconstruction method. 
\section{Analysis of network connectivity under uncertainty}\label{sec2}
\begin{figure*} 
	\centering
	\includegraphics[width=14cm]{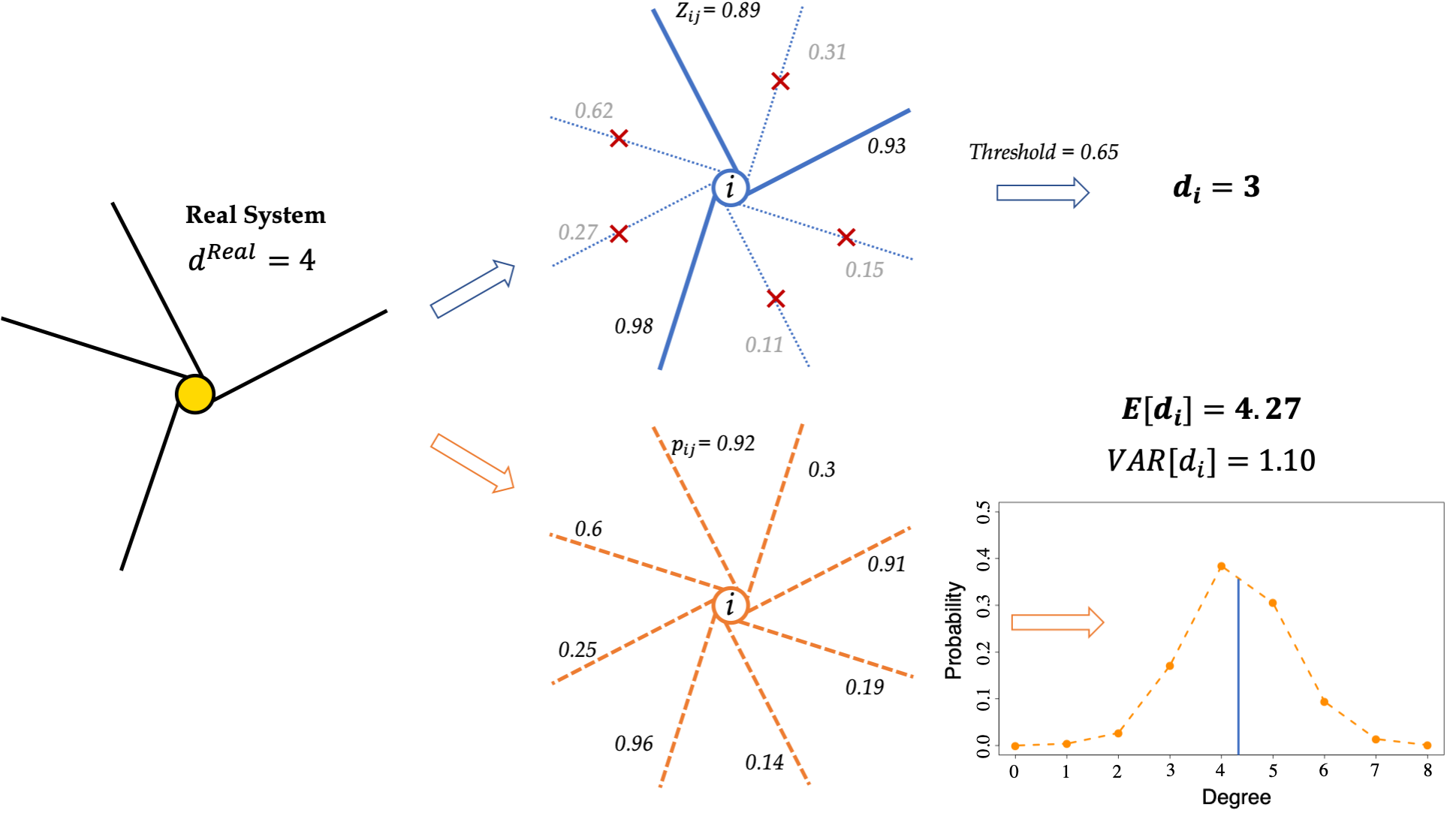}
	\caption{Connectivity reconstruction of a toy system consisting of four edges. Comparison between the widely used thresholding technique (top, blue color), and the redefinition of the node degree as a random variable (bottom, orange color), for an hypothetical real node with degree~4. The node is represented as a node of a ``fuzzy network" (below) in which a probability of existence is associated to each edge. The node degree distribution is plotted on the right-hand side, along with its mean and variance. The expected value results to be closer to the real value than the value of the thresholding process.}\label{fig1}
\end{figure*}
This section describes in detail the process to derive the network descriptors from the observed multivariate time series through the fuzzy network model. 
Given the time series of the nodes' dynamics, a pairwise connectivity measure is computed for each pair of nodes. Subsequently, a bootstrap method is performed for each pair of nodes to derive a \pv for the connectivity. Furthermore, relying on the Bayes theorem, the \pvs are translated into the posterior probabilities of existence of the edges. Consequently, the probabilities are used to define the fuzzy network and the stochastic network descriptors. It is worth remarking that this is only a specific way to obtain a probability for each connection in the system: other approaches, based for instance on inference with explicit generative models \cite{tiago, tiago_sbm} can be used. In fact, the following analysis does not depend on the specific method to obtain probabilities, which are used as input parameters, so that a wider set of problems can be addressed, in which the probabilities $\pi_{ij}$ are directly provided instead of time series. Nevertheless, the study of complex time-varying dynamical systems through time series and connectivity measures has a great explanatory power and significant practical importance. Hence we focused our discussion and numerical experiments on this type of systems. For the sake of simplicity, in what follows we assume the networks to be undirected and unweighted.

\heading{Connectivity Matrix}
The procedures commonly adopted to reconstruct the network topology of a complex system rely on some statistical descriptor used as a proxy for the structural connectivity of the system. These descriptors are able to quantify the relationship between the dynamics of the system's components. In this work we will apply three types of statistical relation: the Pearson correlation coefficient (CC), the Spearman's rank correlation (SC) and the Spectral Coherence (SpeCoh)~\cite{mandel}; an information-theoretic tool: the mutual information (MI)~\cite{shannon}; and a state-space reconstruction tool, namely the Convergent Cross Mapping (CCM)~\cite{sugihara}. These methods have been applied to reconstruct complex networks in different contexts, from neuroscience \cite{bullmore,brain_causality,jeong2001mutual,schiefer2018correlation} to climatology \cite{climate,boers,donges2009backbone,yamasaki}, finance \cite{finance, bonanno2003topology} and ecology~\cite{sugihara}. In general, the problem is to quantify the evidence of the interaction between two components using the information enclosed in the time course of the state vector. The analysis is conducted pairwise, for each pair of components. The result is a matrix which summarizes the strength of the interaction between each pair. We call this matrix \enquote{connectivity matrix} $\mathbf{C}$ to distinguish it from the adjacency matrix
\begin{eqnarray}\label{eq_conmat}
\mathbf{C}= \left[ \begin{array}{ccc} c_{1,1}  & \ldots & c_{1,N}  \\ \vdots  & \ddots & \vdots \\ c_{N,1} & \ldots & c_{N,N}  \end{array} \right].
\end{eqnarray}

\heading{Probabilities of existence}
The statistical significance of the values in the connectivity matrix can be quantified deriving the corresponding \textit{p-values}. To do so, we perform a surrogate data analysis using the reshuffled version of the time series to compute a null model (see Sec.~\ref{numres} for more details), which expresses the null hypothesis $H^0_{ij}$ of lack of connectivity between nodes $i$ and $j$.
The result of this process is a matrix of p-values $p_{ij}$ which quantifies -- for each possible edge $e_{ij}$ -- the strength of the evidence against the null hypothesis $H^0_{ij}$. In the usual reconstruction context the $p_{ij}$ can be used (after adjusting them for the multiplicity) to test against the null hypothesis of lack of connectivity~\cite{nakamura}. These would lead directly to the reconstructed adjacency matrix of the network given a level of significance fixed a priori.

Instead, we ask for the probability that the null hypothesis $H^0_{ij}$ is true, given the $p_{ij}$. That is equivalent to asking for the probability of existence of the edge $e_{ij}$ once the corresponding \pv is known, which reads 
\begin{eqnarray}\label{P(H1)}
P(H^1_{ij}|p_{ij})=1-P(H^0_{ij}|p_{ij})= \pi_{ij}.
\end{eqnarray}
To derive this probability we rely on the work of \cite{sellke} and \cite{held_nomogram}, which provide a Bayesian argument to obtain the posterior probability distribution $P(H^0_{ij}|p_{ij})$ for the null hypothesis $H^0_{ij}$ given the \textit{p-value}~(on the rhs of \eqref{P(H1)}). To determine the functional form of $P(H^0_{ij}|p_{ij})$ from the Bayes theorem, the distribution of the \pvs under the null and alternative hypotheses are needed. It is known that the \pvs under $H^0_{ij}$ are distributed uniformly like $\operatorname{Unif(0,1)}$. This is a direct consequence of the Probability Integral Transform applied to the \pvs~\cite{hung1997behavior}. Instead, under the alternative hypothesis $H^1_{ij}$ the $p_{ij}$ can be considered distributed as a $\operatorname{Beta(\xi,1)}$ probability distribution. This choice reflects the fact that the $p_{ij}$ are bounded between $0$ and $1$ and that under the alternative hypothesis they are skewed on the left (toward $0$).
Since the standard Uniform distribution is a particular case of the Beta distribution ($\xi=1$), it follows that the distribution of $p_{ij}$ is
\begin{eqnarray*}
p_{ij} \sim f(p_{ij}|\xi)=\xi p_{ij}^{\xi-1}
\end{eqnarray*}
so that the parameter $\xi$ includes the information on which hypothesis is considered.
In the Bayesian framework, given a prior distribution $g(\xi)$ for the parameter $\xi$, the test of the null against the alternative hypotheses is assessed by the Bayes factor 
\begin{eqnarray}\label{bf_integral}
B_{g}(p_{ij})=\frac{\operatorname{P}\left(p_{ij} | H^0_{ij}\right)}{\operatorname{P}\left(p_{ij} | H^1_{ij}\right)}=\frac{f(p_{ij} | 1)}{\int_{0}^{1} f(p_{ij} | \xi) g(\xi) d \xi}
\end{eqnarray}
By using the First Mean Value theorem and after some calculations, the inferior Bayes factor is obtained as
\begin{eqnarray}\label{eq_minbf}
\begin{array}{l}
B_{ij}=\inf_{\xi} B_{g}(p_{ij})=\frac{f(p_{ij} | 1)}{\sup _{\xi} \xi p_{ij}^{\xi-1}} \\
\quad=-e p_{ij} \log p_{ij} \quad \text { for } \quad p_{ij}<e^{-1}
\end{array}
\end{eqnarray}
and $B_{ij}=1$ for $p_{ij}>e^{-1}$ where $e$ is the Euler's number. $B_{ij}$ is independent on the parameter $\xi$ and it is valid for any prior distribution on $\xi$. This can be interpreted as a lower bound for the odds of $H^0_{ij}$ on $H^1_{ij}$ given the form of the distribution under $H^1_{ij}$~\cite{sellke}. 
Finally, using the definition, the (inferior) Bayes factor can be mapped into the minimum posterior probability for the null hypothesis given the \pv:
\begin{eqnarray}\label{eq_ppos}
1-\pi_{ij}=\left(1+\left(\frac{B_{ij} \cdot P(H^0_{ij})}{1-P(H^0_{ij})}\right)^{-1}\right)^{-1}
\end{eqnarray}
This formula gives the (maximum) posterior probability $\pi_{ij}$ that the edge $e_{ij}$ exists given the \pv from its connectivity measure, where $P(H^0_{ij})$ is the prior probability for the null hypothesis, which is the only parameter to be fixed in this procedure. This parameter contains the prior knowledge about the possibility of finding an edge between two nodes. In principle, it can assume a different value for every edge in the network, depending on the amount of prior information available at the edge-specific level. In situations where a local characterization of the structure is unavailable, the $P(H^0_{ij})$ can be unique and equal for all the edges, so that $P(H^0_{ij})=P(H^0)$. For instance, a global value can be determined considering information about other networks (e.g. using the expected density of a set of known networks similar to the one under study) or with other problem-specific knowledge; otherwise, an uninformative prior can be used.

\begin{figure*}[!t]
	\centering
	\includegraphics[width=12cm]{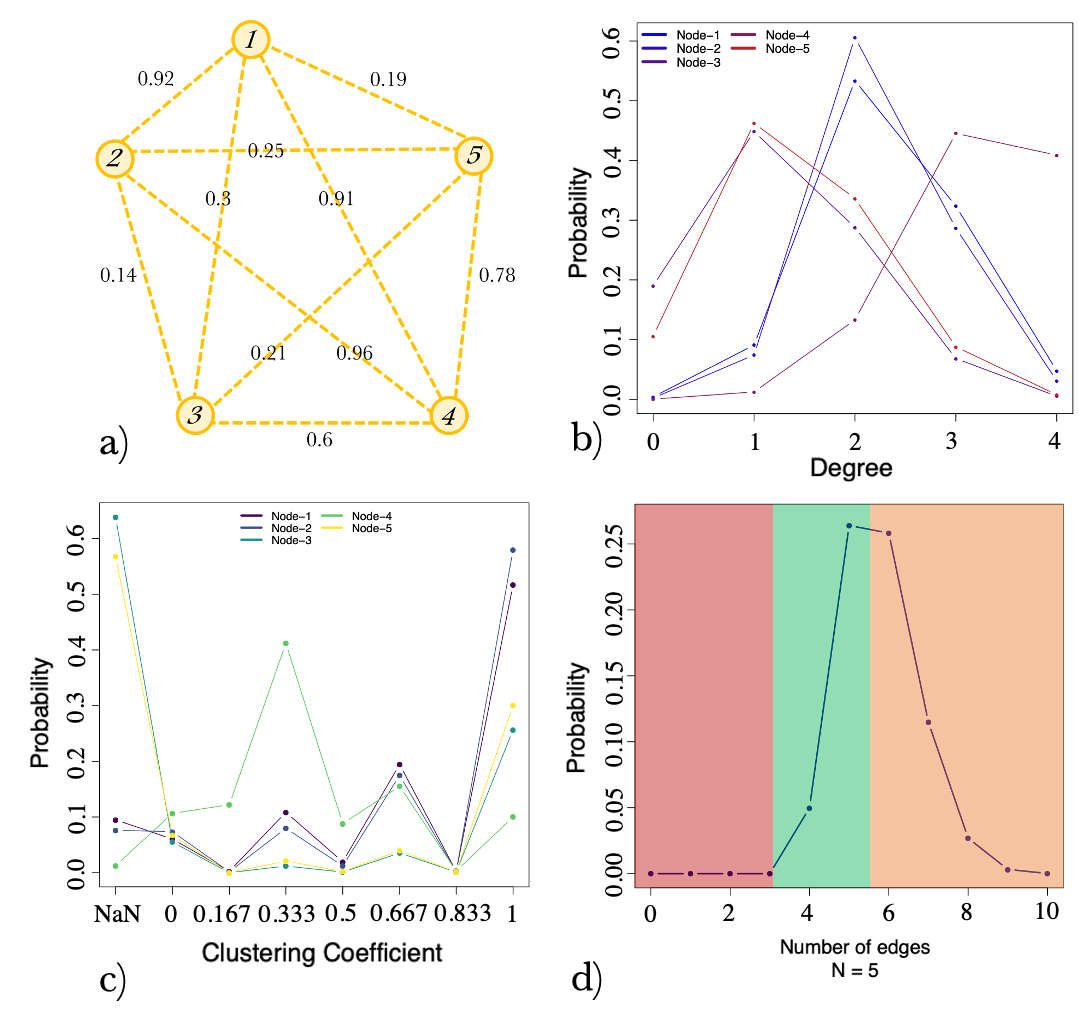}
	\caption{\small{Probability distributions for the degree (b) and the local clustering coefficient (c) for the toy network in (a). The figure (d) represents the probability of having a network of five nodes consisting of a single totally connected component with $i$ edges (x-axis).}\label{fig_wrap}
	}
\end{figure*}

\heading{Building the Fuzzy Network} The probabilities $\pi_{ij}$ of existence of the edge between nodes $i$ and $j$ can be rearranged in a matrix $\bP$, to obtain the probabilistic counterpart of the adjacency matrix:
\begin{eqnarray}\label{eq_fuzzyA}
\bP= \left[ \begin{array}{ccc} \pi_{1,1}  & \ldots & \pi_{1,N}  \\ \vdots  & \ddots & \vdots \\ \pi_{N,1} & \ldots & \pi_{N,N}  \end{array} \right]
\end{eqnarray} 
The matrix $\bP$ resembles  a weighted adjacency matrix, but it has a different meaning: the value $\pi_{ij}$ is not a weight, but it represents the probability of existence of the corresponding edge. Therefore, the matrix $\bP$ totally defines a complete network, whose edges might exist with a certain probability~(see \lfig{fig_wrap}).
This representation, encodes all the knowledge about the structural connectivity of the network. We name this model \enquote{fuzzy network}. 

Given the stochastic nature of the edges, all the structural descriptors must be redefined as random variables. In what follows, we redefine some of the most widely used structural descriptors on the basis of the fuzzy network model.


\heading{Node degree}\label{par:node_degree}
Let us consider a single node in the fuzzy representation of the complex network~(see \lfig{fig1}). The node $i$ has $N$ edges incident to it, each with an associated independent probability of being present. Under this condition, the usual definition of the node degree (i.e. the number of edges incident to the node) is no more applicable, since the node has all the possible degrees at the same time, each with a certain probability. Therefore, another definition of the degree is needed to take into account the uncertainty about the existence of the edges. The most natural choice is to define the degree as a random variable described by its probability distribution, which depends on the probabilities $\pi_{ij}$. The probability that the node $i$ has degree $d_i=k$ can be thought of as the probability to have $k$ successes in a sequence of $N$ independent Bernoulli trials with success probabilities $p_{i1} , p_{i2} ,..., p_{i(N)}$:
\begin{equation}\label{eq_bern}
\begin{array}{ccc}
e_{ij}|\pi_{ij} \sim Bernoulli(\pi_{ij}) &,& d_i = \sum\limits_{j=1}^{N}{[e_{ij}|\pi_{ij}]}
\end{array}
\end{equation}
If the $\pi_{ij}$ were all equal, the probability distribution of the latter sum would be the well-known Binomial distribution. But in this case all the edges incident to node $i$ have different probabilities of existence. 
Consequently, the probability of having $k$ successful trials out of a total of $N$ can be written as
\begin{equation}\label{eq_poibin}
P(d_i=k)=\sum\limits_{{\Lambda\in F_k}}\prod\limits_{j\in \Lambda}{\pi_{ij}\prod\limits_{l\in \Lambda^c}{(1-\pi_{il})}}
\end{equation}
where $F_{k}$ is the set of all subsets of $k$ edges that can be selected from $\{e_{i,1},e_{i,2},e_{i,3},\dots,e_{i,N}\}$. For example, if $N = 3$, then $F_{2}=\left\{\{e_{i,1},e_{i,2}\},\{e_{i,1},e_{i,3}\},\{e_{i,2},e_{i,3}\}\right\}$. $\Lambda^{c}$ is the complement of $\Lambda$, i.e. the set  $\Lambda^{c}=\{e_{i,1},e_{i,2},e_{i,3},... ,e_{i,N}\}\setminus \Lambda$ . This distribution is the so called Poisson-Binomial distribution, and it represents the node degree distribution.
As for the Binomial, the mean is equal to the sum of the $\pi_{ij}$ and the variance is the sum of the probabilities of success times the probabilities of fail:
\begin{equation}\label{eq_musi}
\begin{array}{ccc}
\mu_{d_i}=\sum\limits_{j}{\pi_{ij}} &,& \sigma^2_{d_i}=\sum\limits_{j}{\pi_{ij}(1-\pi_{ij})} 
\end{array}
\end{equation}

Having an entire distribution for each node, we are provided with more information with respect to the case of the usual degree. This additional information makes the calculation of the degree more robust against uncertainty, since it is possible to compute the most significant moments of the distribution.

\heading{Expected degree of a network}\label{par:Ed}
An important summary quantity which characterizes a network is the expected degree of the network. In order to find the expected value for the entire network we exploit the properties of the Poisson-Binomial distribution in \eqref{eq_poibin}. First of all, the Poisson-Binomial distribution is very well approximated by the Normal distribution for fairly small samples (the approximation can also be refined using a continuity correction for discrete random variables). This is a consequence of the Central Limit Theorem (CLT). More precisely, since the Poisson-Binomial is defined as the sum of independent but not identically distributed Bernoulli variables (see \eqref{eq_bern}), the CLT needs to be considered in the Lyapunov formulation, which imposes a condition on the moments of the distribution of $e_{ij}$ in \eqref{eq_bern}~\cite{billingsley}. Suppose ${\{d_1, d_2, . . .d_n\}}$ is a sequence of independent random variables, each with finite expected value $\mu_{d_i}$ and variance $\sigma_{i}$. Let's define $\quad s_{n}^{2}=\sum_{i=1}^{n} \sigma_{d_i}^{2}$. If for some $\delta > 0$, the Lyapunov’s condition 
\begin{eqnarray}
\frac{1}{s_{n}^{2+\delta}} \sum_{i=1}^{n} \mathbb{E}\left[\left|d_{i}-\mu_{d_i}\right|^{2+\delta}\right]=0 
\end{eqnarray}
is satisfied, then the sum of $\frac{d_{i}-\mu_{d_i}}{s_{n}}$ converges in distribution to a standard Normal random variable, as $n$ goes to infinity:
\begin{eqnarray}
\frac{1}{s_{n}} \sum_{i=1}^{n}\left(d_{i}-\mu_{d_i}\right) \stackrel{d}{\rightarrow} \operatorname{Norm}(0,1)
\end{eqnarray}
For the sum of Bernoulli random variables, the Lyapunov condition is easily satisfied (see the \textit{Appendix B} for a detailed discussion) and the convergence is reached even for very small $N$; thus for a network having $N$ nodes, the degree of node $i$ follows:
\begin{eqnarray*}
d_{i} \sim P B\left(\pi_{i, N-1}, \ldots, \pi_{i, N-1}\right) \stackrel{d}{\rightarrow} \operatorname{Norm}\left(\mu_{d_{i}}, \sigma_{d_{i}}\right) 
\end{eqnarray*}
where the parameter $\mu_{d_{i}}$ and $\sigma_{d_{i}}$ are given by \eqref{eq_musi}.
From the properties of the Normal distribution, and from \eqref{eq_musi}, it follows that
\begin{eqnarray}\label{eq_sumdi}
\sum_{i}^{N} d_{i} \approx \operatorname{Norm}\left(\sum_{i}^{N} \mu_{d_{i}}, \sum_{i}^{N} \sigma_{d_{i}}\right)
\end{eqnarray}
In general, the total number of edges $m$  in the network, and the expected degree $c$ can be computed as
\begin{eqnarray}
m=\frac{1}{2} \sum_{i}^{N} d_{i} \quad , \quad c=\frac{2 m}{N}
\end{eqnarray}
which in this case are random variables, since the element $d_i$ is stochastic.
Consequently, we can compute the expected value of the random variable $c$ taking into account \eqref{eq_sumdi}:
\begin{equation}\label{eq_Ec}
\begin{split}
\mathbb{E}[c]=\mathbb{E}\left[\frac{2 m}{N}\right]=\frac{1}{N} \mathbb{E}\left[\sum_{i}^{N} d_{i}\right] \approx \\ \frac{1}{N} \sum_{i}^{N} \mu_{d_{i}}=
\frac{1}{N} \sum_{i}^{N} \sum_{j}^{N} \pi_{ij}
\end{split}
\end{equation}
Therefore, the expected degree for the entire network is twice the sum of the probabilities of existence of all the edges, divided by the number of nodes. This means that picking nodes at random from the network, we expect their degree to be equal to $\mathbb{E}[c]$ (on average) in \eqref{eq_Ec}.

\heading{Clustering Coefficient} The clustering coefficient is defined as the fraction of path of length two that are closed. This coefficient quantifies the transitivity of the network. With transitivity we mean that if node $i$ is connected to node $j$ and node $j$ is connected to node $k$, than also $i$ is connected to $k$~\cite{newman}. This property has fundamental implications on important network characteristics, such as the \enquote{small-worldness}~\cite{watts_strogatz_1998}. The local clustering coefficient is calculated for each node in the network, but other definitions exist for a global measure of the transitivity.  The most common way of defining the local clustering coefficient is the following

\begin{eqnarray}
C=\frac{\# \text { triangles } \times 3}{\# \text { connected triples }}
\end{eqnarray}
where a \enquote{connected triple} is the configuration in which three nodes $ijk$ are connected by the edges $(i, j)$ and $(i, k)$, whereas the edge $(j, k)$ may be present or not. Since each triangle is counted three times when the triples $ijk, jki, kij$ are evaluated, the number of connected triples is divided by $3$.

In the case of a fuzzy network, as for the degree, the clustering coefficient of a node can take all the possible values with a certain probability. Therefore, also this feature must be redefined as a random variable.
The probability of having a certain number of closed triangles in a triple depends on the probabilities of the corresponding edges and on the configurations of the edges in which that number of triangles occurs.
For example, in the fuzzy network of \lfig{fig_wrap}, each node may be tied to a triangle in 6 different ways.
Precisely, in a network of $N$ nodes, each node may be tied to $t$ closed triangles in $\nchoosek{N-1}{t}$ different configurations.
Therefore, the probability of each configuration $c$ can be computed as
\begin{eqnarray}\label{eq_confc}
q_{c}=P\left(\bigcap_{i, j \in S^{c}} e_{i j}\right) \cdot\left[1-P\left(\bigcap_{i, j \in \bar{S}^{c}} e_{i j}\right)\right]
\end{eqnarray}
where $S^c$ is the set of all the pairs of nodes (defining an edge) which define the configuration $c$, and $\bar{S}^{c}$ is its complementary. Since we are assuming that all the edges are independent Bernoulli random variables (\eqref{eq_bern}), the intersection in \eqref{eq_confc} can be taken out of the parentheses and replaced with the summation.
The configurations can be considered as mutually disjoint and collectively exhaustive events, so that~$\sum_{c} q_{c}=1$. Consequently, the set of all configurations is regarded as the sample space of the network reconstruction experiment.
In conclusion, the clustering coefficient probability distribution for the node $i$ is given by
\begin{eqnarray}\label{eq_clus}
P_{i}^{cc}(C=\widetilde{C})=\bigcup_{c \in \Gamma_{i}^{\tilde{C}}} q_{c}=\sum_{c \in \Gamma_{i}^{\tilde{c}}} q_{c}
\end{eqnarray}
where $\Gamma_{i}^{\tilde{C}}$ is the set of the configurations in which the node $i$ has clustering coefficient equal to $\tilde{C}$. A representative example of the distribution is shown in \lfig{fig_wrap}\textit{b}, which shows the clustering coefficient distribution for each node of the depicted toy network.


\heading{Connected Components}
Another fundamental feature of a complex network is the existence of a global connected component. A global connected component exists if there is at least one path from any node to any other node. Again, this feature is subjected to the stochasticity of the edges. The objective is to find the probability that all the nodes of the network belong to a unique connected component of $k$ edges. To find the probability this we need to label the configurations which make the network completely connected with the related probability. Therefore, we can employ again the \eqref{eq_confc} also to address the problem of the connectivity. In particular, we are asking for the probability that a network of $N$ nodes, is completely connected by $k$ edges. This probability reads
\begin{equation}
P_{k}^{cn}=\bigcup_{c \in \Gamma_{\mathbf{k}}^{cn}} q_{c}=\sum_{c \in \Gamma_{\mathbf{k}}^{cn}} q_{c}
\end{equation}
where $q_c$ comes from \eqref{eq_confc} and $\Gamma_{\mathbf{k}}^{cn}$ is the set of the configurations in which exactly $k$ edges make the network connected. These configurations can be efficiently found with a Breadth-First algorithm. The union sign on the left can be replaced by the summation because, as mentioned before, the set of the configurations is the sample space of the experiment, so that all the configurations are disjointed.  An example is shown in \lfig{fig_wrap}\textit{c}, which illustrates the probability that the five nodes of the toy network belong to a single totally connected component with $k$ existing edges (x-axis). Specifically, in the red window there are not enough edges to connect the network; in general, to do so are necessary at least $N-1$ edges. In the green window the probability increases and reaches the maximum. Finally, increasing further the number of edges required to connect the network, the probability decreases~(orange window). It seems counter-intuitive that the probability of having a connected network decreases increasing the number of edges; the reason is that above a certain number of edges, the entire configuration becomes less likely, since the probability $q_c$ of existence of all the $k$ edges (at once) is smaller.


\section{Numerical experiments and results}\label{numres}
This section reports the results of the fuzzy network analysis for a set of synthetic and real-world networks. In both cases we considered undirected and unweighted networks. In particular, we used 15 different synthetic network structures: 5 Erd{\H o}s-Rényi, 5 Barabasi-Albert, 5 Watts-Strogatz, with 256 nodes each. The parameters of the generative models were fixed so that all the networks have expected degree approximately equal to 12. The three real-world networks considered are the collaboration network between Jazz musician~\cite{arenas_jazz}, the food web of Little Rock Lake~\cite{little_rock} and the brain network of the \textit{Rhesus macaque}~\cite{macaque}. For each of the resulting 18 structures we generated 5 different dynamical realizations of two dynamical models: a linear (auto-regressive moving-average) ARMA(5,3) 
\begin{eqnarray}
x_{t}=\sum_{i=1}^{5} \alpha_{i} x_{t-i}+\sum_{i=1}^{3} \beta_{i} \varepsilon_{t-i}+\varepsilon_{t}+\gamma
\end{eqnarray}
where $\alpha_{1},\ldots ,\alpha_{5}$ and $\beta_{1},\ldots ,\beta_{3}$ are the model parameters for the auto-regressive and moving-average parts respectively, $\varepsilon_{t}$ is a white noise random variable and $\gamma$ is a constant;
and a non-linear logistic model~\cite{may1976}
\begin{eqnarray}
x_{n+1}=r x_{n}\left(1-x_{n}\right)
\end{eqnarray}
where the parameter $r$ is chosen randomly for each model realization in the interval $[3.57,3.82]$ to assure a chaotic regime.
Each time series spans a time horizon of 1024 time-steps. All the models take into account the connectivity of the underlying network using linear coupling terms, which are chosen to be small enough to guarantee that the resulting time series, especially in the case of the ARMA model, remain stationary. The result is a total of 180 numerical experiments.
 
The data used for the subsequent analysis are the time courses of the state variable of the nodes. Starting from this information we computed the connectivity matrix for all the networks using all the methods mentioned in Sec.~\ref{sec2}, obtaining a value of connectivity $c_{ij}$ for each pair of nodes $(i,j)$. The statistical significance of the connectivity was assessed by computing the corresponding \textit{p-values} obtained by means of surrogate data analysis. Specifically, an adequate null hypothesis $H^0_{ij}$ is the lack of relationships between the nodes $i$ and $j$, which can be easily achieved by reshuffling the observed time course at each site~\cite{lancaster}. The reshuffled time series possess the same mean, variance, and histogram distribution as the original signal, but any temporal correlation is destroyed, making this null model adequate to test for coherence or causal relations between nodes'~dynamics. Nevertheless, other types of surrogates techniques can be used depending on the null hypothesis one would like to test. For instance, if one is interested in testing against the null hypothesis that the time series are correlated like in a random linear process, one should opt for Iterated Amplitude-Adjusted Fourier Transform-based surrogates, which preserve the linear features of the time series even in the frequency domain, while washing out higher-order dependencies. There are many additional types of surrogates that one can use to test against other null models, but the optimal choice is beyond the scope of the present paper. Changing null hypothesis could change the probability distributions of nodes' descriptors accordingly, but this neither negatively affects the goodness of the proposed method nor it can be easily related to the sensitivity of the method.
The alternative hypothesis $H^1_{ij}$ is that such relationships exist. If $c^0_{ij}$ is the value of connectivity expected by chance for the edge $e_{ij}$, $H^0_{ij}$ corresponds to $c^{obs}_{ij}=c^0_{ij}$, while $H^1_{ij}$ is $c^{obs}_{ij} \neq c^{obs}_{ij}$. Given the empirical distribution of $c^0_{ij}$ it was possible to obtain the \pv $p_{ij}$ corresponding to the value of connectivity $c_{ij}$ from the original time series. Subsequently, the resulting \pvs were used to obtain the minimum posterior probabilities through \eqref{eq_minbf} and \eqref{eq_ppos}. As mentioned in Sec.~\ref{sec2}, the only free parameter of the process is the prior probability for the null hypothesis $P(H^0_{ij})$. In this experiment, we set it equal to $1-D$ for all the edges, where $D$ is the average density of the networks considered. With this choice we are allowed to write the prior probability as $P(H^0)$, without the subscript ${ij}$. This is clearly not the optimal choice for the prior probability, since it is not true that it is equivalent for all the edges. However, this puts us in a scenario where only a global prior information about the network structure is available. The computation of the minimum posterior probability for all the $p_{ij}$ returns the adjacency matrix $\mathbf{P}$ of the fuzzy network, which is used to compute all the network descriptors as described in the previous section. It is to be noticed that, in general, the prior $P(H^0_{ij})$ value determines a shift in the probability $\pi_{ij}$ of $\mathbf{P}$ in the interval $[0,1]$; the same happens in the particular case in which $P(H^0_{ij})=P(H^0)$.

\begin{figure}
\centering
\begin{subfigure}[t]{0.4\textwidth}\label{fig_distribution_a}
\centering
	\includegraphics[width=6.9cm]{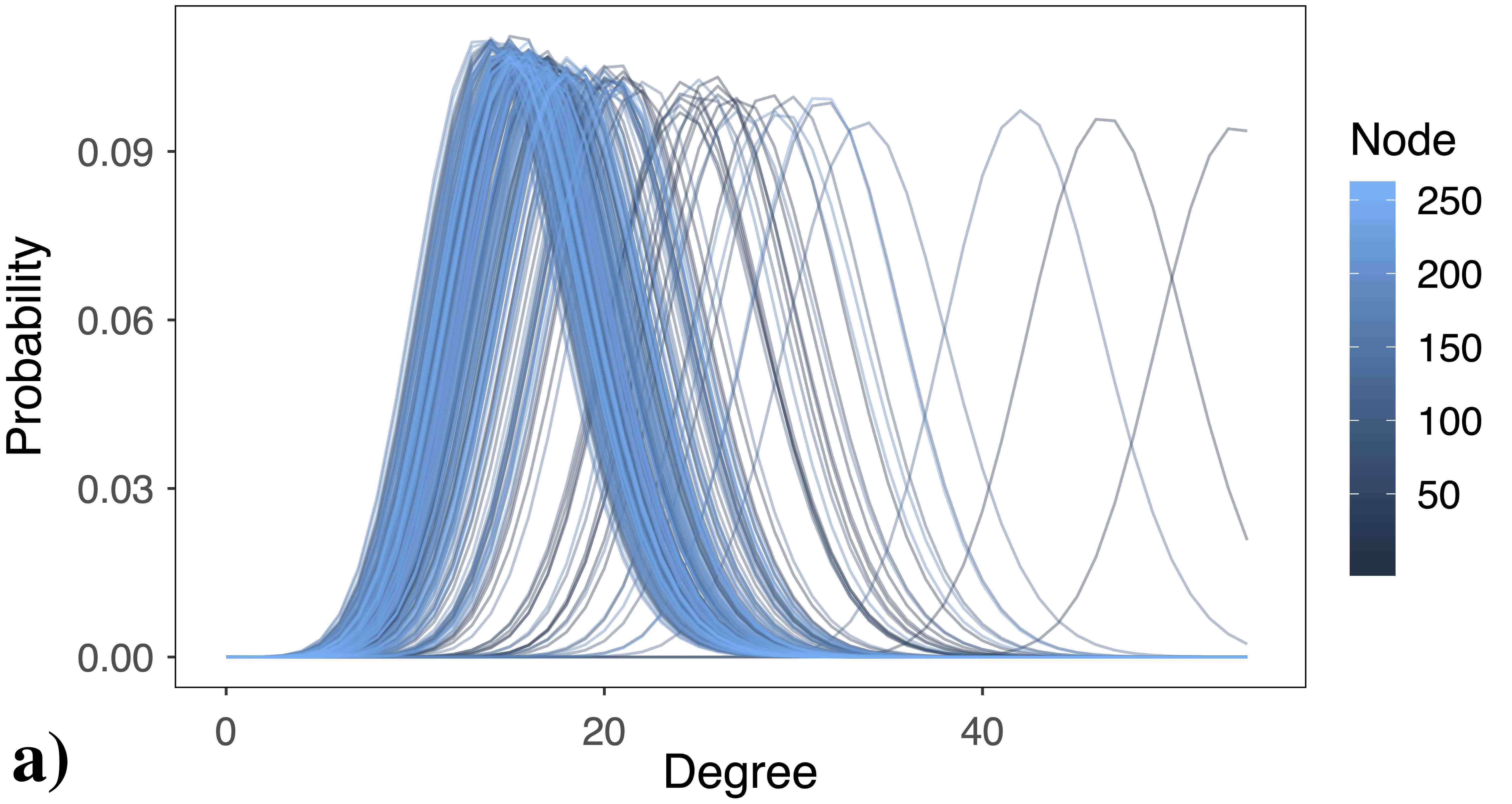}
	\bigskip
\end{subfigure}
\begin{subfigure}[t]{0.4\textwidth}\label{fig_distribution_b}
\centering
	\includegraphics[width=6.9cm]{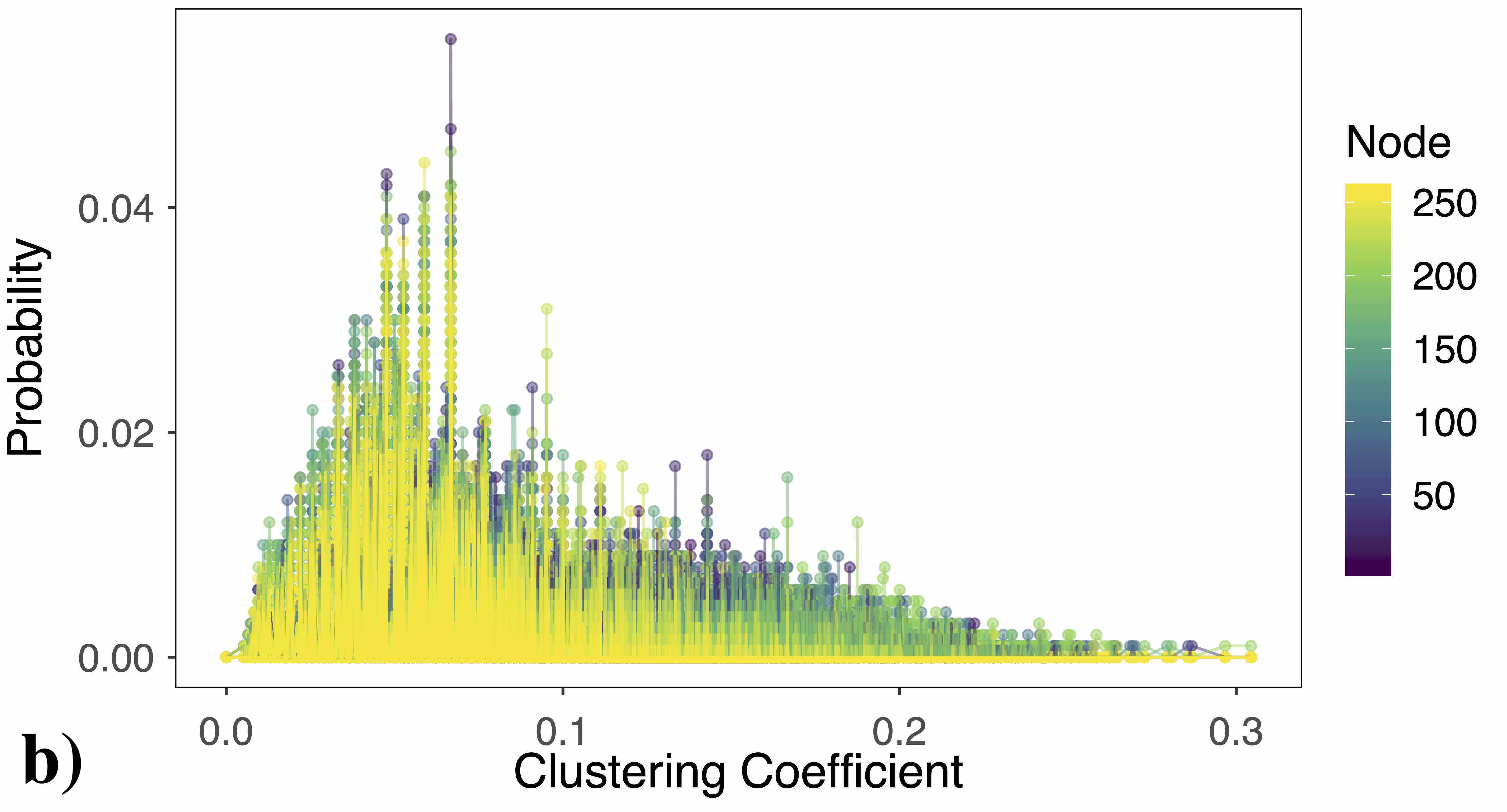}
	\bigskip
\end{subfigure}
\begin{subfigure}[t]{0.4\textwidth}\label{fig_distribution_c}
\centering
	\includegraphics[width=6.9cm]{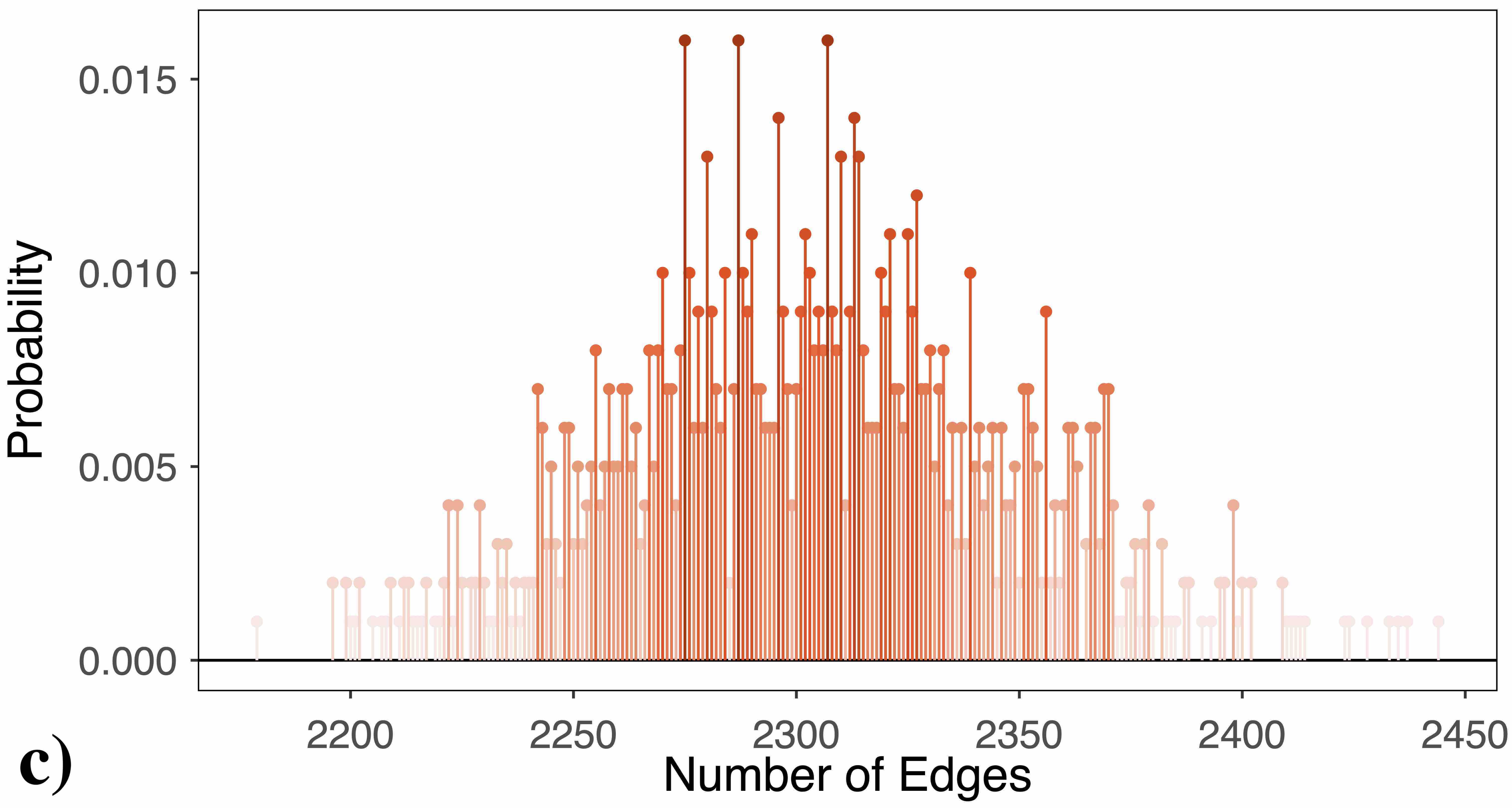}
	\medskip
\end{subfigure}
\begin{minipage}[t]{.4\textwidth}
\centering
\caption{A single realization of the ARMA dynamics on a single realization of the Barabasi-Albert network. Probability mass functions for the node degree (a), for the local clustering coefficient of the nodes (b) and for the connected component (c) as obtained from the fuzzy network analysis introduced in this study. }\label{fig_distributions}
\end{minipage}
\end{figure}

Unlike the traditional methods for network reconstruction, which use heuristics to determine a threshold on $c_{ij}$ or on $p_{ij}$, our process maintains the uncertainty on the parameters until the computation of the network descriptors.
Figure~\ref{fig_distributions} shows an example for the realization 1 of the Barabasi-Albert network, using the Pearson correlation coefficient with ARMA dynamics. The three figures are analogous to those in \lfig{fig_wrap} for the toy network. The \lfig{fig_distributions}a) shows the probability mass functions for the degree of each node, which follow the Poisson-Binomial distribution in \eqref{eq_poibin}. We used a free \textsf{R} routine for the numeric approximation of the analytical distribution provided by~\cite{hong_poibin}. It is to be noticed that the vast majority of the distributions are grouped around the real average degree of the network, that is $12.8$. Our theoretical median prediction in this case is $15.03$ with $[11,21]$ $68\%$ confidence interval.

Similarly, \lfig{fig_distributions}b) shows the probability mass functions for the local clustering coefficient of the nodes. The distributions are very irregular and do not follow any known probability function. For comparison, the average clustering coefficient in the real network is $0.107$  whereas our theoretical median prediction is $0.0603$ with $[0.0521,0.1116]$ $68\%$ confidence interval.

Finally, in \lfig{fig_distributions}c) is reported the probability mass function for the connectivity, which represents the probability that all the nodes of the network belong to the same connected component of $k$ edges. The distribution overestimates the number of edges needed to have a unique connected component, since the ground-truth network, which is actually connected in one component, consists of 1515 edges. A possible explanation for this overestimation is that the $\pi_{ij}$ might be too high due to the choice of the prior $P(H^0)$ in \eqref{eq_ppos}, whose effect is to shift the values on the vertical axis.

A useful synthesis of these network descriptors (besides the expected value) is the maximum posterior probability (MPP) estimate, which is simply the mode of the computed distributions. Furthermore, other statistics about the dispersion can be computed to assess the uncertainty on the values at node level. This is made possible by the fuzzy approach which does not require any threshold – neither on the connectivity matrix nor on the \pvs – allowing the uncertainty to be considered as part of the network analysis, rather than an obstacle to overcome.

\begin{figure*} 
	\centering
	\includegraphics[width=\textwidth]{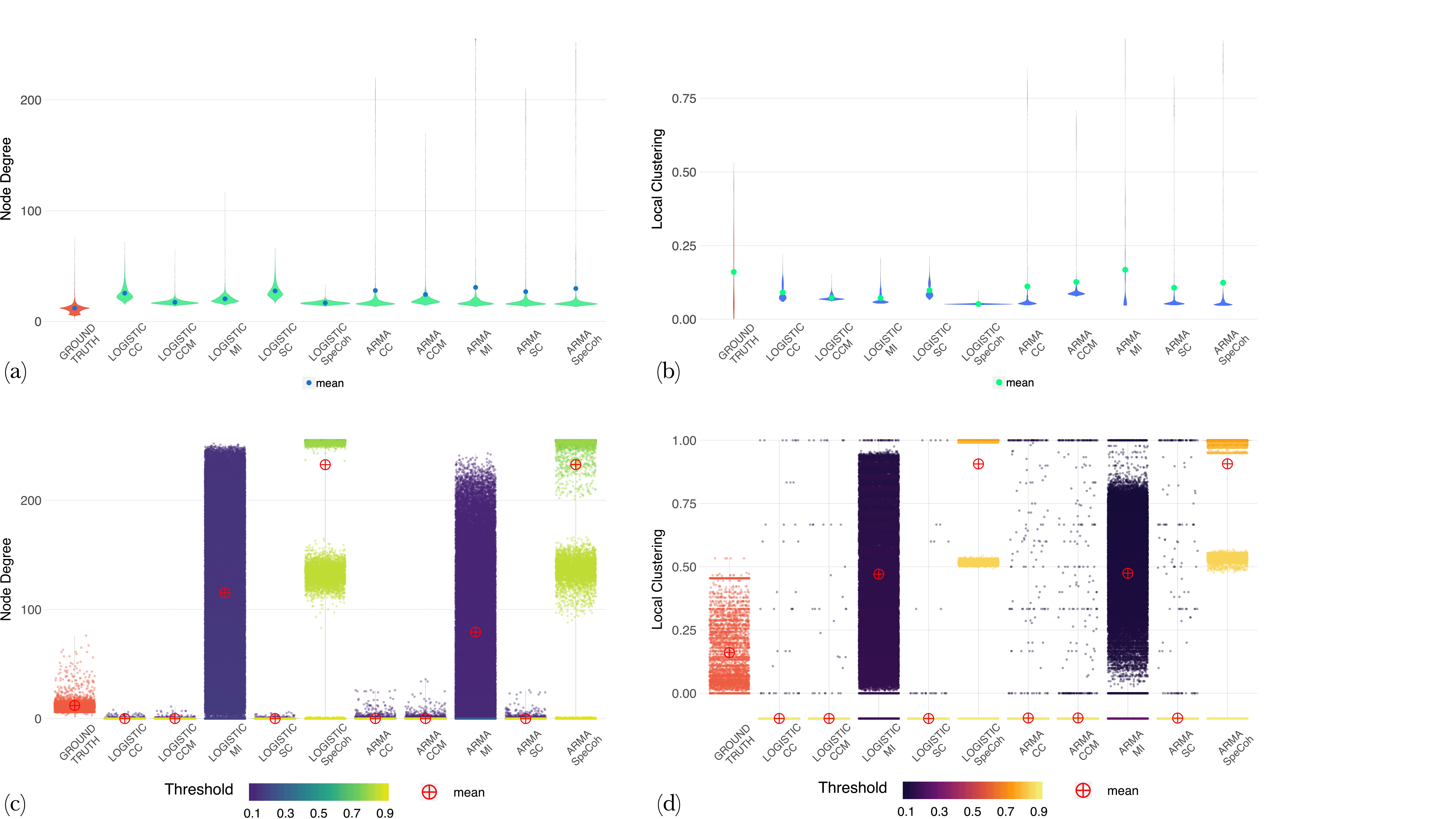}
	\caption{Summary distribution for the degree (a) and the local clustering coefficient (b) for all the synthetic network, grouped by dynamics and connectivity measures. The figures (c) and (d) show the distributions of the degree and the local clustering coefficient by varying the threshold level, which is encoded by the color, and with respect to different dynamics (ARMA or LOGISTIC) and discriminating statistics for network reconstruction (CC, CCM, MI, SC, SpeCoh; see the text for details). Each point corresponds to a single node (a random horizontal jitter is added): note that the same node appears multiple times with different colors, for each value of the threshold. In every plot it is also indicated the mean of the distributions.}\label{fig_dc}
\end{figure*}

\begin{figure*} 
	\centering
	\includegraphics[width=\textwidth]{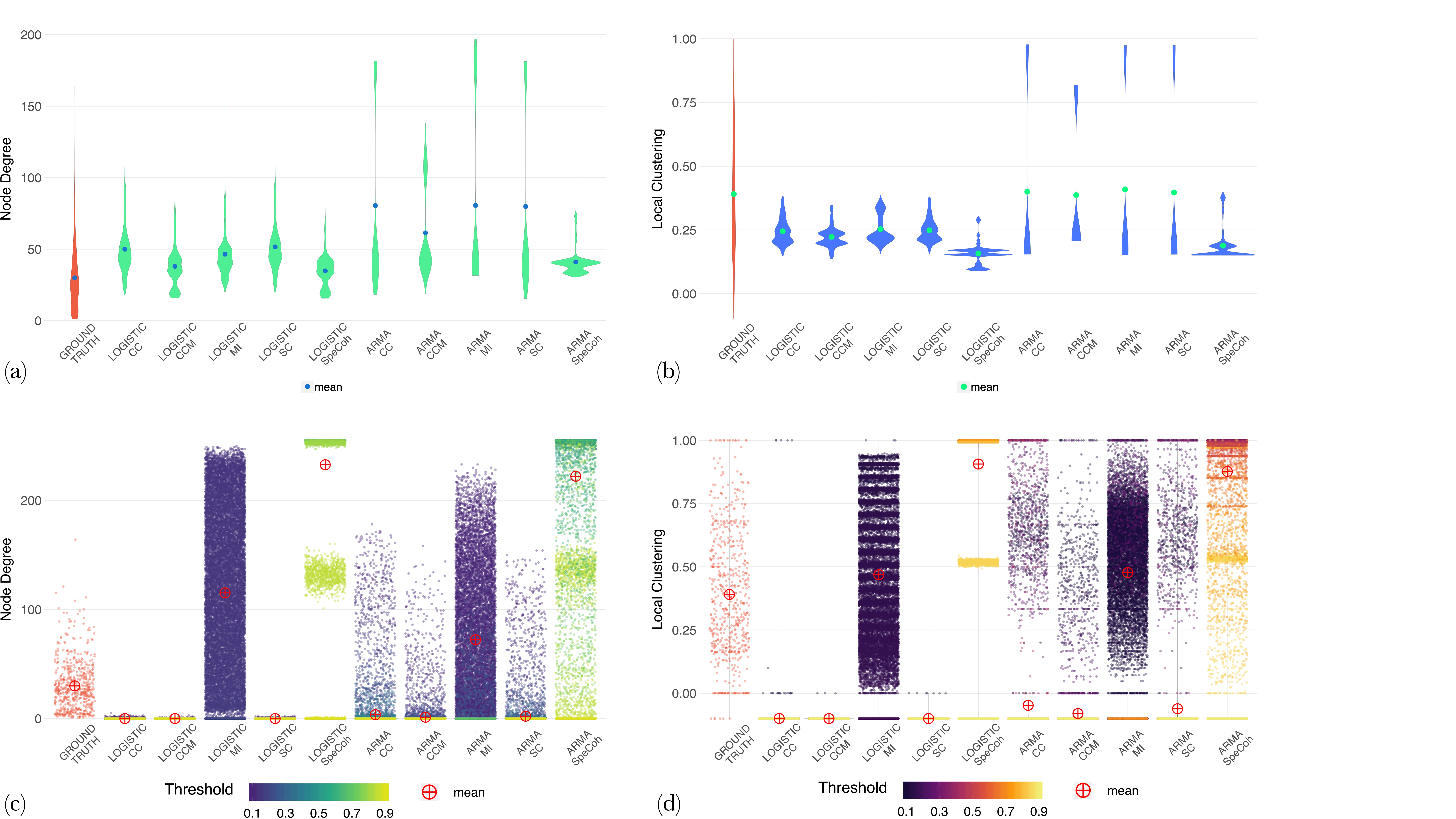}
	\caption{Summary distribution for the degree (a) and the local clustering coefficient (b) for the real-world network, grouped by dynamics and connectivity measures. The figures (c) and (d) show the distributions of the degree and the local clustering coefficient by varying the threshold level, which is encoded by the color, and with respect to different dynamics (ARMA or LOGISTIC) and discriminating statistics for network reconstruction (CC, CCM, MI, SC, SpeCoh; see the text for details). Each point corresponds to a single node (a random horizontal jitter is added): note that the same node appears multiple times with different colors, for each value of the threshold. In every plot it is also indicated the mean of the distributions.}\label{fig_rdc}
\end{figure*}

Figures~\ref{fig_dc} and \ref{fig_rdc} show an overview on the results of the whole set of numerical experiments. The figures include the comparisons between the fuzzy procedure and another well-known method, which consists of thresholding the connectivity matrix $\mathbf{C}$ (\eqref{eq_conmat}) to derive the binary adjacency matrix.
Several criteria exist to fix a value for the threshold, (see e.g. \cite{threshold_choice_methods, fallani_optimal_threshold}). Instead, for the sake of comparison, we reconstructed the networks with several threshold levels, according to the range of the connectivity provided by each tool. In particular we used equally spaced values in the interval $[-1;1]$ for the statistical tools (CC, SC, SpeCoh) and for the Convergent Cross Mapping (CCM), whereas equally spaced quantiles were supplied for the Mutual Information (MI).
The plots in Fig.~\ref{fig_dc}a--d  are obtained from the aggregating the 150 synthetic network experiments, while those in Fig.~\ref{fig_rdc}a--d are aggregated over the 30 real network experiments. The results are grouped according to the dynamic and the discriminating statistics used to assess the connectivity. 
Each point in Fig.~\ref{fig_dc}c--d and \ref{fig_rdc}c-d represents the value of degree and clustering for a single node given a threshold value for the connectivity matrix.

Overall, analysis shows that the majority of the results obtained from the fuzzy analysis are consistent with the expectations, whereas the thresholding approach, regardless of the statistical method, tends to underestimate or overestimate the true values for varying thresholds. These results provide a strong indication that results from threshold models not only strongly depend on the value of the threshold, but also that there can be no thresholds for which, on average, reliable measure of network indicators as simple as degree centrality and local clustering coefficient can be obtained.

Let us discuss in greater detail the results concerning the fuzzy network analysis. All the discriminating statistics yield meaningful results in terms of expected value although, in the considered cases our method slightly overestimates the average degree. The clustering coefficients instead, show better average values for the linear dynamics, despite a general slight underestimation. Both types of deviation from expected values of the features are due to the shift effect of~$P(H^0)$.
In this specific setting, the lower values of the two descriptors are not well captured, because every edge has a positive – albeit very low – probability of existence which keeps the degree and the clustering away from zero. 

As expected, the fuzzy network analysis applied to ARMA dynamics returns the less valuable outcomes, arguably because of the very low coupling that we imposed to the edges. This result is still consistent with our expectations: a low value of the coupling results in statistical correlations more difficult to detect even in the case of linear dynamics. In this case, because of this additional source of uncertainty due to such a limitation, the inferred values span a broader interval for both degree and clustering coefficient: nevertheless, most of the mass meets the ground-truth distribution, suggesting that the fuzzy network analysis is able to robustly cope with the increased level of uncertainty. 
The best tool to derive the connectivity proved to be the CCM, which allows the non-linearity to be taken into account adequately. 

It is worth noting that the performance of statistical methods and the overall results may be improved by adjusting for the spurious relationships such as the partial correlations, but a direct implementation of this task is beyond the scope of the present work.
 
Remarkably, the fuzzy descriptors outperform the traditional thresholding reconstruction methods in all the cases. Despite the numerous threshold levels in place, the real values of the network features are rarely detected by the latter reconstruction technique. In some cases, the results reflect the ground-truth, but only for specific values of the threshold which remain basically arbitrary.

The results for the real-world networks are qualitatively analogous (Fig.~\ref{fig_rdc}), whereas the performances for the ARMA models have improved for both the network descriptors. The real-world networks span a broader range of values for both the node degree and the clustering coefficient. This is clearly reflected in both the methods presented. Even in this case, the distribution of the two network descriptors are skewed towards the lower values; this feature is mostly captured by the fuzzy model, while just specific values of the threshold accomplish the ground-truth.

\section{Future directions}

Here we provide a brief discussion about future directions. In fact, this work opens the way to the definition of other descriptors -- e.g. centrality measures -- of complex systems in the wake of the fuzzy descriptors. Also, the fuzzy perspective might be extended to the dynamical features of a complex network by studying, for instance, the properties of the fuzzy counterpart of the Laplacian. 

A first example of another topological descriptor eligible to be redefined in fuzzy terms is the rich-club coefficient~(see e.g. \cite{richclub}). Using the information coming from the degree distribution for the individual nodes, the rich-club coefficient can be redefined as the probability $P^{rich}(e,n,k)$ to observe $e$ edges connecting $n$ nodes of degree greater than $k$. Having the Poisson-Binomial probability distribution for the degree (\eqref{eq_poibin}) and the probabilities of the possible configurations (\eqref{eq_confc}), all the ingredients are there to obtain $P^{rich}$. Given a fuzzy network of $N$ nodes and the associated fuzzy adjacency matrix with elements $\pi_{ij}$ the probability that the node $i$ has degree greater than $k$ is 
\begin{equation}
    P^0=P(d_i>k) = 1-CDF(\mathfrak{P}[k,\mathbf{\mathbf{\pi}}_{i\cdot}])
\end{equation}
where $j \in {1,...,N}$ and $j\neq i$ and $\mathfrak{P}[k,\pi_{ij}]$ is the Poisson-Binomial distribution of parameters $\overrightarrow{\mathbf{\pi}}_{i\cdot}$ (the row $i$ of the fuzzy adjacency matrix). Let's select $n$ nodes with $n \in {2,3,...,N}$ from the possible $\sum\limits_{n=1}^\infty{ {N}\choose{n}}$ configurations. The configurations are indexed as $c$  with $c\in [1, {{N}\choose{n}}]$. The nodes selected in the particular configuration $c$ form a set $S_c$ with cardinality $|S_c|=n$. The probability that \textit{all} the $n$ nodes $i \in S_c$ of the configuration $c$ have degree greater than $k$ is 
\begin{equation}
P^{>k}_c=P(d_{S_c}>k)=\prod\limits_{i\in Sc}{P^0_c}=\prod\limits_{i\in Sc}{P(d_i>k)}
\end{equation}
under the hypothesis of independence. Finally, the probability of existence of $e$ edges between the nodes of degree greater than $k$ is given by
\begin{equation}
    P^{rich}_c=P(E_{>k}=e)=P^{>k}_c \cdot \mathfrak{P}[e,\overrightarrow{\mathbf{\pi}}_{c}]
\end{equation}
where $\overrightarrow{\mathbf{\pi}}_{c}$ is the vector of probabilities of existence of the edges in the configuration $c$. This probability makes use again of the Poisson-Binomial distribution as it is capable to model the presence of the edges in the configurations. To obtain the probability distribution $P^{rich}(e,n,k)$ for the rich-club coefficient of the network, regardless of the configuration, the inclusion-exclusion criterion must be applied to the above equation.

Leveraging on the descriptors showed so far, it is possible to compute the probability of observing a random walk occupying a particular node.
The probability is usually given by $\frac{k}{2m}$: in the case of a fuzzy network it is replaced by the ratio distribution of a Poisson-Binomial and a Gaussian distribution~(see \eqref{eq_poibin} and \eqref{eq_sumdi}). 

The information about the degree distribution can also be exploited to study the fuzzy counterpart of assortativity of the network, for example starting by defining the excess degree distribution for the single node and consequently for the entire network. 

We expect that further analysis in this direction, left for future studies, will lead to interesting results from both theoretical and applied perspectives.

\section{Conclusion}

In this work we have presented a novel framework for network analysis under uncertainty about the underlying connectivity, which overcomes some of the issues typical of network reconstruction procedures. This framework can be used to infer the structural features of a complex system when its topological connectivity is specified by edge probabilities.  Also, we proposed a simple method, mathematically grounded, for the computation of these probabilities from a set of p-values, usually obtained from one's preferred analytical technique.

The leading idea is to define a new standpoint, from which the uncertainty about the structural features of a complex system can be detected and quantified, without the explicit reconstruction of the underlying network structure. The method does not require any assumption on the topology of the network under study, nor on the underlying generative process.
Conversely, our approach enables one to include prior knowledge about the existence of the single edges in a rigorous manner, without fixing any arbitrary threshold nor level of statistical significance. Starting from the \pvs associated with measures of correlations or statistical causality, we obtained a Bayesian definition of the probability of existence of the single edge. The probabilities are rearranged in an adjacency matrix, which represents a \enquote{fuzzy} model used to elicit relevant information on the network structure. All the information have a stochastic nature which allows uncertainty to be assessed. Consequently, we proposed new definitions of some important network descriptors such as the node degree and clustering coefficient, considering them as random variables.
Finally, we compared the results with a very well-known method for network reconstruction, showing the strength and weakness of our procedure. 
From a computational perspective, for small networks the probabilities of the configurations can be directly computed with \eqref{eq_confc}, which gives all the information to derive the probabilities for the clustering coefficient and the connectivity. In case of large networks the computational effort might be prohibitive: however an adequate sampling procedure from the fuzzy adjacency matrix in \eqref{eq_fuzzyA} can be performed to compute the distributions and the statistics of interest.

The method may be enhanced by applying different tools which take into account the partial relations in deriving the connectivity matrix. In addition, past studies can be readily integrated with our approach given the \pvs obtained therein.

\bibliographystyle{apsrev4-1}
\bibliography{biblio.bib}

\section*{APPENDIX A: FROM P-VALUE TO PROBABILITY}

\begin{figure}[ht]
    \centering
    \includegraphics[width=8cm]{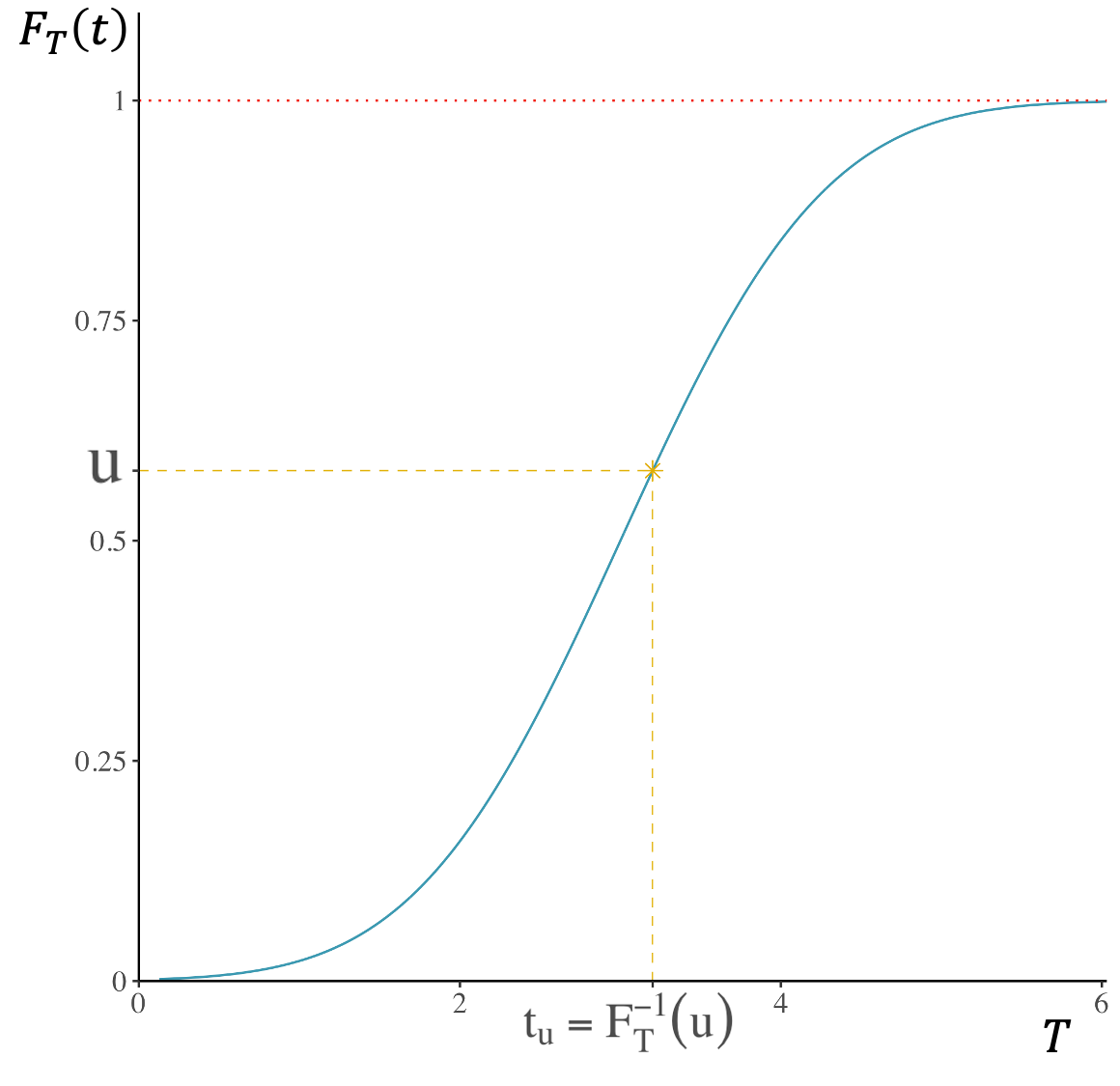}
    \caption{\small{Probability Integral Transform example}\label{pit_plot}
    }

    \begin{minipage}[!h]{.4\textwidth}
        \centering
    \end{minipage}
\end{figure}

In general, the \textit{p-values} can be formally considered as random variables (see e.g.~\cite{murdoch2008p}), which under the null hypothesis are distributed uniformly in $[0,1]$. This is a direct consequence of the Probability Integral Transform: given the test statistic $T$ of interest, and its realization $t$, the \textit{p-value} is by definition (using the same notation as in the manuscript)
\begin{eqnarray*}
    \begin{aligned}
        p_{ij}=P\left(T \geq t \mid H^0_{ij}\right) &= 1-P\left(T<t \mid H^0_{ij}\right) \\
        &=1-F^0_{T}(t)
    \end{aligned}
\end{eqnarray*}
All the subsequent formula are intended under the null hypothesis, so we drop the $0$ at the apex. 
Let's now define the random variable $U=F_T(t)$~(see also \textit{Fig.\ref{pit_plot}}). It follows that
\begin{eqnarray*}
    \begin{aligned}
        F_T(t)=P\left(U \leq u \right) &= P\left(F_T(t_u) \leq u \right) = P\left(T \leq t_u \right) \\
        &= P\left(T \leq F^{-1}_T(u) \right) = F_{T}\left(F_T^{-1} \left(u\right)\right) \\
        & = u
    \end{aligned}
\end{eqnarray*}
which is equivalent to the definition of a Uniform distribution for the variable $U$.
Since $p_{ij} = 1-F_T(t)=1-U$ we need to prove that also $F_P(p_{ij}) \sim \operatorname{Unif}(0,1) $:

\begin{eqnarray*}
    \begin{aligned}
        F_P\left(p_{ij}\right)=P\left(P \leq p \right) &= P\left(1-U \leq p_{ij} \right) \\
        &= P\left(U \geq 1-p_{ij} \right) \\
        &= 1-P\left(U \leq 1-p \right) \\
        &= 1-1-p\\
        &= p \hspace{0.5cm} \square
    \end{aligned}
\end{eqnarray*}

It is to be noticed that, under the alternative hypothesis, the \textit{p-values} are not uniformly distributed, but are typically skewed. Therefore, as we explained in the manuscript, the distribution of the \textit{p-value} $p_{ij}$ (for the edge $e_{ij}$) is modeled as $\operatorname{Beta(\xi,1)}$ following the procedure of \cite{sellke}:
\begin{eqnarray}\label{beta}
p_{ij} \sim f(p_{ij}|\xi)=\xi p_{ij}^{\xi-1}.
\end{eqnarray}
The standard Uniform distribution is a particular case of \eqref{beta}, where the parameter $\xi=1$
\begin{eqnarray*}
\operatorname{P}\left(p_{ij} | H^0_{ij}\right) = f(p_{ij}|\xi=1)=1
\end{eqnarray*}
The \eqref{bf_integral}, in the main text, reports the definition of the Bayes Factor, from which we want to prove that a lower bound for the odds of $H^0_{ij}$ on $H^1_{ij}$ is represented by \eqref{eq_minbf}, synthetically reported here:
\begin{eqnarray*}\label{sellke}
B_{ij} = \inf_{\xi} B_{g}(p_{ij}) = -e p_{ij} \log p_{ij} \quad \text { for } \quad p_{ij}<e^{-1}
\end{eqnarray*}
and equal to 1 otherwise. 

Since the numerator of \eqref{bf_integral} is a constant, the lower bound for $B_{ij}$ corresponds to the upper bound of the denominator:
\begin{eqnarray}\label{supdenom}
B_{ij}=\inf_{\xi} B_{g}(p_{ij})=\frac{1}{\sup _{\xi} \int_{0}^{1} f(p_{ij} | \xi) g(\xi) d \xi}
\end{eqnarray}
From the First Mean Value Theorem we know that in general, if $f:[a, b] \rightarrow R$ is continuous and $g$ is integrable and does not change sign on $[a, b],$ then there exists some $c$ in $(a, b)$ such that
\begin{eqnarray*}
\int_{a}^{b} f(x) g(x) d x=f(c) \int_{a}^{b} g(x) d x
\end{eqnarray*}
In our case we have that
\begin{eqnarray*}
    \begin{aligned}
        \int_{0}^{1} f\left(p_{i j} \mid \xi\right) g(\xi) d \xi &= \int_{0}^{1} \xi p_{i j}^{\xi-1} \cdot g(\xi) d \xi \\
        &= \bar{\xi} p_{i j}^{\bar{\xi}-1} \int_{0}^{1} g(\xi) d \xi \\
        &=\bar{\xi} p_{i j}^{\bar{\xi}-1}
    \end{aligned}
\end{eqnarray*}
Thus we can rewrite the \eqref{supdenom} as 
\begin{eqnarray}\label{bf_fmvt}
B_{ij}=\inf _{\xi} B_{g}\left(p_{ij}\right)=\frac{1}{\operatorname{sup}_{\bar{\xi}}\bar{\xi} p_{ij}^{\bar{\xi}-1}}
\end{eqnarray}
We now define $h \left(\bar{\xi}\right)=\bar{\xi} p_{i j}^{\bar{\xi} -1}$, so that
\begin{gather*}
h^{\prime}(\xi)=p^{\bar{\xi}-1}+\bar{\xi}\left(p^{\bar{\xi}-1} \ln p\right)=0 \\
p^{\bar{\xi}-1}=-\bar{\xi} \ln (p) \cdot p^{\bar{\xi}-1} 
\end{gather*}
which is true only if $\bar{\xi}=-\frac{1}{\ln p}$. Substituting this result in \eqref{bf_fmvt} we obtain
\begin{eqnarray*}
    \begin{aligned}
        B_{ij}=\frac{1}{-\frac{1}{\ln p} \cdot p^{-\left(\frac{1}{\ln p}+1 \right)}} &= -\ln p \cdot p^{\frac{1}{\ln p}} \cdot p \\
        &= -p \ln p \cdot p^{\log _{p} e} \\
        &= - e p \ln p \hspace{0.5cm} \square
    \end{aligned}
\end{eqnarray*}
Note that, since for $p>e^{-1}$ the function $B_{ij}$ is decreasing, we impose that for \textit{p-values} larger than $e^{-1} \approx 0.368 \rightarrow B_{ij}=1$.

Given the definition of the Bayes Factor as 
\begin{eqnarray*}
B_{g}\left(p_{i j}\right)=\frac{\mathrm{P}\left(p_{i j} \mid H_{i j}^{0}\right)}{\mathrm{P}\left(p_{i j} \mid H_{i j}^{1}\right)}=\frac{\mathrm{P}\left( H_{i j}^{0} \mid p_{i j} \right)\cdot \mathrm{P}\left( H_{i j}^{1}\right)} {\mathrm{P}\left(H_{i j}^{1} \mid p_{i j} \right) \cdot \mathrm{P}\left( H_{i j}^{0}\right) }
\end{eqnarray*}
it follows that 
\begin{eqnarray*}
    \begin{aligned}
        \mathrm{P}\left(H_{i j}^{0} \mid p_{i j}\right) &= B_{i j} \cdot \frac{\mathrm{P}\left(H_{i j}^{0}\right) \mathrm{P}\left(H_{ij}^{1} \mid p_{i j}\right)}{1-\mathrm{P}\left(H_{i j}^{0}\right)} \\
        &=B_{i j} \cdot \frac{\mathrm{P}\left(H_{i j}^{0}\right) \left( 1-\mathrm{P}\left(H_{i j}^{0} \mid p_{ij}\right)\right)}{1-\mathrm{P}\left(H_{i j}^{0}\right)}
    \end{aligned}
\end{eqnarray*}
which finally gives
\begin{eqnarray*}
1-\pi_{ij}=\left(1+\left(\frac{B_{ij} \cdot P(H^0_{ij})}{1-P(H^0_{ij})}\right)^{-1}\right)^{-1}
\end{eqnarray*}
which is the \eqref{eq_ppos} of the main text.

\section*{APPENDIX B: LYAPUNOV CONDITION FOR THE POISSON-BINOMIAL DISTRIBUTION}

In section \ref{par:node_degree} we defined the fuzzy degree of a single node, which is a random variable following the Poisson-Binomial distribution. In order to obtain the expected degree of a network we relied on the fact that the Poisson-Binomial converges to the Normal distribution if the Lyapunov condition were satisfied. Here we prove that the condition is satisfied under in very broad conditions.

Let $d_i \sim$ Bernoulli $\left(p_{i}\right),$ with $d_{1}, d_{2}, \ldots$ independent but not identically distributed random variables, represent the degree of node $i$ as stated in \eqref{eq_bern}.
Let also $X_{i}=d_i-\mu_{d_i} = d_i-p_{i}$. Defining $\quad s_{n}^{2}=\sum_{i=1}^{n} \sigma_{d_i}^{2}$ we can rewrite the Lyapunov condition as
$$
\lim _{n \rightarrow \infty} \frac{1}{s_{n}^{2+\delta}} \sum_{i=1}^{n} \mathbb{E}\left[\left|X_i\right|^{2+\delta}\right] = 0 \\ 
\Longrightarrow \frac{1}{s_{n}}\sum_{i=1}^{n} X_{i} \stackrel{d}{\rightarrow} N(0,1).
$$
We prove that the Poisson-Binomial probability distribution satisfies this condition, by finding an upper bound converging to zero to the above sum. 

To do so we observe that
\begin{eqnarray*}
    1 \geq p_{i}\left(1-p_{i}\right)= \sigma_{d_i}^2 = \mathbb{E}\left[X_i^{2}\right] \geq \mathbb{E}\left[\left|X_i\right|^{2+\delta}\right]
\end{eqnarray*}
for any $\delta>0$. Therefore,
$$
\begin{aligned}
    & \frac{1}{s_{n}^{2+\delta}} \sum_{i=1}^{n} \mathbb{E}\left[\left|X_i\right|^{2+\delta}\right] \leq \frac{1}{s_{n}^{2+\delta}} \sum_{i=1}^{n} \mathbb{E}\left[X_i^2\right] \\ 
    &= \frac{1}{s_{n}^{2+\delta}} \sum_{i=1}^{n} \sigma_{d_i}^2
    =\frac{1}{s_{n}^{\delta}}
\end{aligned}
$$
Consequently, since $s_{n} \rightarrow \infty$ (exept for degenerate cases where $p_i=0$ or $p_i=1$ for all $i$), the Lyapunov condition is satisfied and similarly it is the ``normalized'' Poisson-Binomial random variable follows $\sum_{i=1}^{n} X_i / s_{n} \stackrel{d}{\rightarrow} N(0,1)$.

\section*{}

\section{Acknowledgments}
The authors thank Alice~Schwarze and Jean-Gabriel~Young for useful discussions and valuable suggestions.

\end{document}